\documentclass[aps,prl,floatfix,twocolumn]{revtex4}
\usepackage{graphicx,amsmath,units,xspace}

\graphicspath{{figures/}}

\newcommand{\micron}{\ensuremath{\unit{\mu m}}\xspace}

\begin{document}

\title{Diffusion of single ellipsoids under quasi-2D confinements}

\author{Y. Han,$^{1,2}$ A. Alsayed,$^{2}$ M. Nobili,$^{3}$ and A. G. Yodh$^{2}$}
 \affiliation{$^{1}$ Hong Kong University of Science and
 Technology, Clear Water Bay, Hong Kong, China}
 \affiliation{$^{2}$ Department of Physics and Astronomy, University of Pennsylvania\\
   209 South 33rd St., Philadelphia, PA 19104 USA}
\affiliation{$^{3}$Laboratoire des Colloides, Verres et
Nanomateriaux CNRS-University Montpellier II, Place E. Bataillon
34090 Montpellier France}
\date{\today}

\begin{abstract}
We report video-microscopy measurements of the translational and
rotational Brownian motions of isolated ellipsoidal particles in
quasi-two-dimensional sample cells of increasing thickness.   The
long-time diffusion coefficients were measured along the long
($D_a$) and short ($D_b$) ellipsoid axes, respectively, and the
ratio, $D_a/D_b$, was determined as a function of wall confinement
and particle aspect ratio.  In three-dimensions this ratio
($D_a/D_b$) cannot be larger than two, but wall confinement was
found to substantially alter diffusion anisotropy and substantially
slow particle diffusion along the short axis.
\end{abstract}

\maketitle

\section{Introduction}

In many biological and industrial processes, diffusing particles are
non-spherical and move in confined geometries.  Examples of
particles in this scenario include proteins diffusing in membranes
\cite{Saffman75} and very fine grains migrating through pores in
porous media. To date, quantitative measurements of anisotropic
particle diffusion in confined geometries have been limited.
However, new particle fabrication and imaging technologies combined
with new image analysis tools now make the direct measurement of the
diffusion of anisotropic particles readily possible.  Thus, in this
contribution we investigate the anisotropic diffusion of isolated
ellipsoidal particles confined between two parallel plates.

The Brownian diffusion coefficient $D$ of an isolated spherical
particle is well understood.  It is inversely proportional to the
drag (or friction) coefficient $\gamma$ via the Einstein relation,
\begin{equation}
D=k_BT/\gamma \label{eq:Einstein}
\end{equation}
where $k_B$ is the Boltzmann constant and $T$ is the temperature.
For a prolate spheroid with long axis of length $2a$ and two short
axes of length $2b$, translational diffusion is anisotropic and is
described by diffusion coefficients $D_a=k_BT/\gamma_a$ along the
long axis, and $D_b=k_BT/\gamma_b$ along the short axes. The
rotational diffusion coefficient of the prolate spheroid about its
short axes is $D_{\theta}=k_BT/\gamma_{\theta}$.  Generally, the
drag coefficients $\gamma_a$, $\gamma_b$ and $\gamma_{\theta}$
depend on the shape and size of the ellipsoid.  Brownian motion of
anisotropic particles was first seriously considered by F. Perrin
\cite{Perrin34,Perrin36} who computed these drag coefficients
analytically for a spheroid diffusing in three dimensions (3D).
Interestingly, the ratio $D_a/D_b$ varies from one to two in 3D, as
the spheroid aspect ratio $\phi=a/b$ varies from one to infinity.

The problem of diffusion in confined geometries, such as quasi-2D
media, is different from the 3D case as a result of a complex
interplay between hydrodynamic drag,  the boundaries of the medium,
and the particle geometry. Surfaces near a moving particle modify
fluid flow fields, often increasing particle hydrodynamic drag. A
full theoretical formulation of wall hydrodynamic effects has been
developed for one sphere (or ellipsoid) coupled to one wall
\cite{Happel91}. However, for more complicated situations, such as a
sphere or an ellipsoid confined by two parallel walls, the only
available analytical solutions are for weak confinement in a few
special symmetric configurations \cite{Happel91}.  Recent numerical
calculations \cite{Bhattacharya05}, on the other hand, have been
developed to derive the hydrodynamic drag of a single sphere and a
linear chain of spheres confined more strongly in quasi-2D.

On the experimental side, the hydrodynamic drag of single spheres in
weak confinement have been measured \cite{Lin00,Dufresne01}, and
video microscopy has been applied recently to measure anisotropic
particle diffusion, including ellipsoids in quasi-2D \cite{Han06}
and 3D \cite{Mukhija07}, colloidal clusters near one wall
\cite{Kim08}, and carbon nanotubes in weak confinement
\cite{Bhaduri08}.  In the present contribution we report
measurements of hydrodynamic drag on ellipsoids in quasi-2D,
confined between two parallel walls.  We explore the strong
confinement regime where drag coefficients are not readily available
from theory and simulation, and we report on a light interference
method to accurately measure the confinement.  We find that the
diffusion anisotropy is made stronger and the diffusion along
ellipsoid short axes is dramatically slowed due to wall confinement.
The experiment and analyses are similar to a previous paper
\cite{Han06}.  However the scope of the present work is different,
focusing instead on how confinement affects diffusion, rather than
on the detailed time-dependent Brownian dynamics of a single
ellipsoid with the greatest diffusion anisotropy.

\section{Theory background}

When a spheroid with semi-axes ($a$, $b$, $b$) moves along one of
its principle axes with velocity $v$, through an unbounded quiescent
fluid with viscosity $\eta$ at low Reynolds number, then the
translational and rotational (about short axis) drag coefficients
affecting the spheroid are
\begin{subequations}
\label{eq:Stokes}
\begin{equation}
\gamma =6\pi \eta b G,
\end{equation}
\begin{equation}
\gamma_{\theta}=6 \eta V G_{\theta}.
\end{equation}
\end{subequations}
$V$ is the volume of the spheroid and $G$ is the geometric factor
that renders the ellipsoid different relative to the case of a
sphere. The geometric factors for prolate spheroids diffusing in 3D
are analytically given by Perrin's equations \cite{Happel91}:
\begin{subequations}
\label{eq:G}
\begin{equation}
G_{a} = \frac{8}{3}\frac{1}{\left[
\frac{2\phi}{1-\phi^2}+\frac{2\phi^2-1}
{(\phi^2-1)^{3/2}}\ln(\frac{\phi+\sqrt{\phi^2-1}}{\phi-\sqrt{\phi^2-1}}) \right]},\\
\label{eq:Ga}
\end{equation}
\begin{equation}
G_{b} = \frac{8}{3}\frac{1}{\left[
\frac{\phi}{\phi^2-1}+\frac{2\phi^2-3}
{(\phi^2-1)^{3/2}}\ln(\phi+\sqrt{\phi^2-1}) \right]},\\
\label{eq:Gb}
\end{equation}
and~\cite{Koenig75,Perrin34}
\begin{equation}
G_{\theta} = \frac{2}{3}\frac{\phi^4-1}{\phi \left[ \frac{2\phi^2-1}
{\sqrt{\phi^2-1}} \ln(\phi+\sqrt{\phi^2-1}) - \phi \right]}.\\
\label{eq:Gtheta}
\end{equation}
\end{subequations}

\begin{figure}[!t]
  \centering
  \includegraphics[width=0.8\columnwidth]{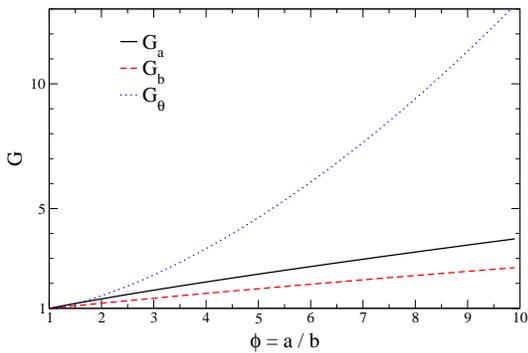}
  \caption{The geometric factors $G$ in Eq.~(\ref{eq:G}) as a function of aspect ratio $\phi$.}
  \label{fig:G}
\end{figure}

Here $\phi=a/b$ is the aspect ratio. When $\phi=1$, then
$G=G_{\theta}=1$, and Eq.~(\ref{eq:Stokes}) reduces to the
translational and rotational Stokes laws for a sphere. Note also
that Eqs.~(\ref{eq:Stokes}) and (\ref{eq:G}) are obtained using
stick boundary conditions, valid when the particle is much larger
than fluid molecules \cite{Hu74,Bauer74}. In Fig.~\ref{fig:G},
Eq.~(\ref{eq:G}) is plotted out as a function of $\phi$ for $\phi$
less than 10.  When the aspect ratio $\phi \gg 1$,
Eqs.~(\ref{eq:Einstein}), (\ref{eq:Stokes}) and (\ref{eq:G}) yield
\begin{equation}
D_a=\frac{k_B T\ln\phi}{2\pi \eta a},
 \qquad D_b=\frac{k_B T\ln\phi}{4\pi \eta a}.
\end{equation}
The ratio between these diffusion coefficients along long and short
axes, i.e. $D_a/D_b=G_b/G_a$, increases monotonically from one to
two as $\phi$ increases from one to infinity (in 3D).  In quasi-2D,
however, $D_a/D_b$ can be larger than two.

\section{Experiment}

The diffusion of micrometer size PMMA (polymethyl methacrylate) and
PS (polystyrene) ellipsoids was measured in water confined between
two glass walls. Both PS and PMMA ellipsoids are synthesized by the
method described in Ref.~\cite{Ho93}. Briefly, we placed 0.5\% (by
weight) PS spheres into a 12\% (by weight) aqueous PVA (polyvinyl
alcohol) solution residing in a Petri dish. After water evaporation,
the PVA film was stretched at 130$^\circ$C. The PS (or PMMA) spheres
embedded in the film are readily stretched because their glass
transition temperatures are below 130$^\circ$C. After cooling to
room temperature, the PVA was dissolved and ellipsoids obtained.
Note, the initial PMMA or PS spheres must not be cross-linked,
otherwise they cannot be stretched.   We measured the size of
ellipsoids by SEM and by optical microscopy.

The ellipsoid solutions were cleaned and stabilized with 7 mM SDS
(sodium dodecyl sulfate). The ellipsoids were not expected to have
strong interactions with the glass surfaces, because the solution
ionic strength was more than 0.1 mM and the Debye screening length
for the particles was correspondingly less than 30 nm. However, it
is difficult to estimate the ionic strength accurately in a thin
cell because the glass surfaces can release Na$^+$ ions
\cite{Crocker96}. Nevertheless, we found that the addition of 2 mM
salt to the solution did not induce a detectable change in particle
diffusion coefficients. This observation suggests that the double
layers are not significantly affecting particle diffusion.

Glass surfaces of the sample cell were rigorously cleaned in a 1:4
mixture of hydrogen peroxide and sulfuric acid by sonication. Then
the glass was thoroughly rinsed in deionized water and quickly dried
with an air blow gun. Typically 0.3 $\unit{\mu L}$ solution spread
over the entire $1.8\times 1.8~\textrm{mm}^2$ coverslip area, and
ellipsoids did not stick to the surfaces. Because the gravitational
height, $k_BT/mg$, is much larger than the cell thickness, $H$, the
ellipsoids were readily suspended around mid-plane between the two
walls. Finally, the cell was sealed with UV cured adhesive (Norland
63).

\begin{figure}[hbt]
  \centering
  \includegraphics[width=\columnwidth]{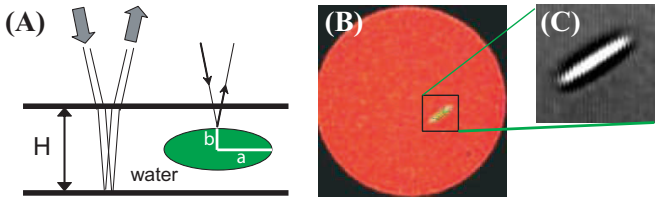}
  \caption{(color online) (A) Schematic of sample dimensions and the interference observation mode.  (B) True interference color image of ellipsoid in the reflection mode of the microscope. (C) Bright field ellipsoid image in the transmission mode.}
  \label{fig:microimages}
\end{figure}

We measure the wall separation by light interference. When the cell
thickness is below a few micrometers, then the interference colors
produced by reflections from the two inner surfaces of the sample
walls in white light illumination can be observed by eye or in the
reflection mode of microscope, see Fig.~\ref{fig:microimages}A,B.
When the wall separation $H=0$, the effective light path difference
is $\Delta l=\lambda/2$ due to the $\pi$ phase shift of reflection
at the lower surface. Thus all wavelength components of the white
light yield a dark black color in interference at $H=0$. When $H>0$,
the reflection light in the normal direction is a mixture of light
with various wavelengths, and different wavelengths contribute with
different weights to the observed color. White light interference
from a wedge, for example, will be bands of colors as in the
Michel-Levy Chart \cite{Hartshorne70}. By comparing the observed
color with the Michel-Levy Chart, we can effectively read out the
corresponding $\Delta l$ and obtain $H=\Delta l/(2 n_w)$, where
$n_w$ is the refractive index of water. In the Michel-Levy Chart,
the color starts from black at $\Delta l=0$ and changes from red to
blue periodically with period $\Delta l=625$ nm. To avoid misreading
the color by one or more periods, we either made a reference wedge
or we put dilute spacer spheres with known diameter between the
glass slides to establish a reference thickness. Also, color bands
may shift slightly because the illumination light is not an ideal
white light source.  This error however, should be less than
$625/4~\textrm{nm}$, so that the error of $H$ is less than $\delta H
~(625$ nm$/4)/(2n_w)=60$ nm.  Although the absolute value of $H$ may
be subject to $\sim 60~\unit{nm}$ uncertainty as described above,
the relative values of different $H$ in one cell should be more
accurate ($\sim 30~\unit{nm}$) because we can easily distinguish
more than 8 different colors in one band including deep red, light
red, orange, light orange, yellow etc.

Usually our sample thickness had less than 20 nm variation in the
central $1~\textrm{mm}^2$ area and had 1-2 $\mu$m variation over the
whole $18\times18~\textrm{mm}^2$ area. Thus, we can study the
diffusion of ellipsoids at different $H$ in one cell. The
interference between reflected light from top inner surface of the
wall and the ellipsoid's top surface give rise to different colors
(see Fig.~\ref{fig:microimages}). As is the case with Newton's
rings, the interference colors due to the two ellipsoid tips and the
center of the ellipsoid were different.  We found that the color
only fluctuated near the two ellipsoid tips; the color was quite
constant near the ellipsoid center.  Thus the height fluctuation of
the ellipsoids was very small and tumbling motions in the vertical
plane were not strong.

Particle motions observed by microscopy were recorded by a CCD
camera to videotape at 30 frames/sec. In the dilute suspension, only
one ellipsoid was visible in the $640 \times 480~\textrm{pixel}^2 =
38.4\times 51.2~\mu \textrm{m}^2$ field of view under 100$\times$
objective during a half-hour experiment. We defocus slightly so that
the ellipsoid can be more accurately located along its long axis.
The built-in 2D Gaussian fit function in IDL (Interactive Data
Language) was used to locate the center and orientation of the
ellipse in each video frame. In practice, a small percent
($\sim$3\%) of the frames failed to be correctly tracked. Without
these frames, the trajectory breaks into short pieces and very
long-time behavior becomes difficult to measure. To capture these
frames, we very slightly adjusted tracking parameters or image
contrast and re-analyzed the images; after these corrections roughly
$3\% \times 3\% = 0.09\%$ of the frames remain incorrectly tracked.
We then repeated this procedure iteratively until all $\sim$50000
frames in one dataset were correctly tracked. The mean square
displacements (MSD) at time lag $t=0$ has small non-zero intercept
due to the tracking errors. Thus we can estimate the spatial and
angular resolution from intercepts of their corresponding MSDs
\cite{Crocker96a}. The orientation resolution is $1^{\circ}$, and
spatial resolutions are 0.5 pixel = 40 nm along the particle's short
axis and 0.8 pixel = 64 nm along its long axis because of the
superimposed small tumbling motion.

\begin{figure}[hbt]
  \centering
  \includegraphics[width=0.8\columnwidth]{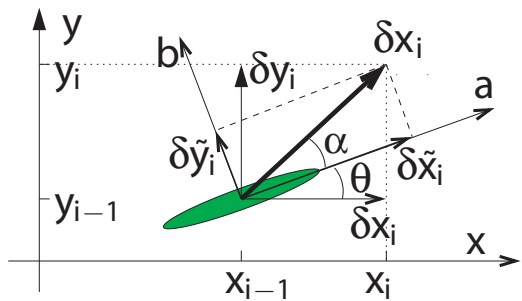}
  \caption{An ellipsoid in the $x$-$y$ lab frame
            and the $\tilde{x}$-$\tilde{y}$ body frame. The angle
            between two frames is $\theta(t)$. The displacement
            $\delta \mathbf{x}$ can be decomposed as ($\delta \tilde{x}$,
            $\delta \tilde{y}$) or ($\delta x$, $\delta y$).}
  \label{fig:coordinates}
\end{figure}

From the image analysis, we obtained the trajectory of a particle's
center-of-mass positions $\mathbf{x}(t_n) = (x(t_n), y(t_n))$ in the
lab frame and its orientation angle $\theta(t_n)$ relative to the
$x$-axis at times $t_n = n \times (1/30)$ sec, see
Fig.~\ref{fig:coordinates}. We define each $1/30$-sec time interval
as a step. During the $n^{\rm th}$ step, the particle's position
changes by $\delta \mathbf{x} (t_n) = \mathbf{x}(t_n ) -
\mathbf{x}(t_{n-1} )$ and its angle by $\delta \theta (t_n) = \theta
(t_n ) - \theta ( t_{n-1})$. To obtain the drag coefficients along
long and short axes, we need to covert the measured displacements
from the fixed lab frame to the local body frame. Step displacements
$\delta \tilde{\mathbf{x}}_n$ relative to the local body-frame and
step displacements $\delta \mathbf{x}_n$ relative to the fixed lab
frame are related via \label{eq:projection}
\begin{equation}
\left(
 \begin{array}{c}
 \delta \tilde{x}_n \\ \delta \tilde{y}_n
 \end{array}
\right) = \left(
 \begin{array}{cc}
   \cos \theta_n & \sin \theta_n \\
   -\sin\theta_n & \cos \theta_n
 \end{array}
\right) \left(
 \begin{array}{c}
   \delta x_n \\ \delta y_n
 \end{array}
\right),
\end{equation}
where $\theta_n=(\theta(t_{n-1})+\theta(t_n))/2$, see
Fig.~\ref{fig:coordinates}. In practice, choosing
$\theta_n=\theta(t_{n-1})$ or $\theta_n=\theta(t_n)$ has little
effect on our results because $\theta$ barely changes during 1/30 s.

\begin{figure}[!t]
  \centering
  \includegraphics[width=\columnwidth]{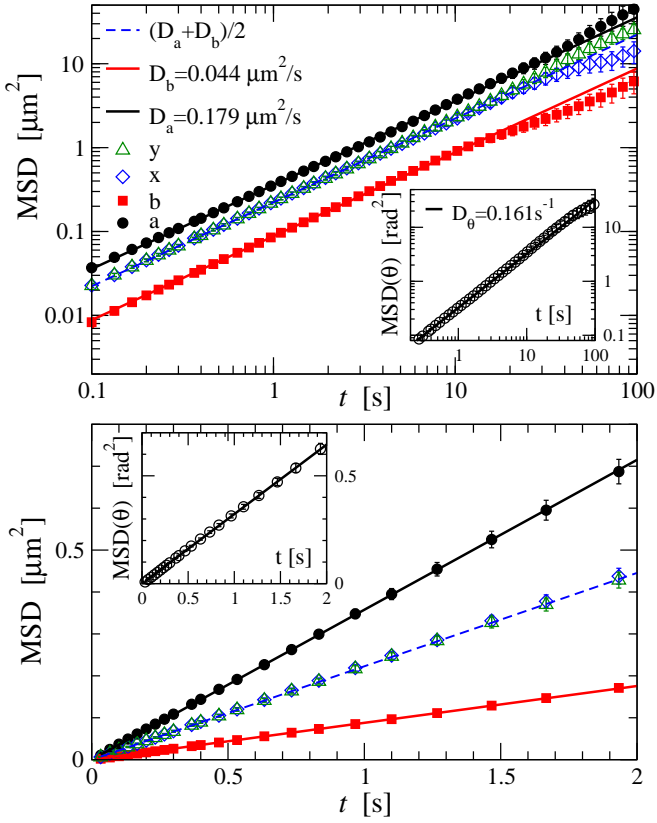}
  \caption{(color online) Mean square displacements (MSDs) of a
$2.4\times 0.3 \times 0.3~\micron^3$ ellipsoid confined in an
$846~\unit{nm}$ thick cell. Top panel:
  log-log plot. Bottom panel: linear plot. The four lines arranged from top to bottom, respectively, are MSDs along $a$,
$x$, $y$ and $b$ axes. Symbols represent experimental data, and
lines represent linear fits to these data.  Insets: Angular MSDs.
All curves exhibit diffusive behavior, and the diffusion
coefficients, $D=\textrm{MSD}/(2t)$, shown in the figure are derived
from the best fit lines.}
  \label{fig:MSD}
\end{figure}

Figure~\ref{fig:MSD} shows mean-square-displacements (MSDs) of a
$2.4\times 0.3 \times 0.3~\micron^3$ ellipsoid confined in an
$846~\unit{nm}$ thick cell. In both the lab and the body frame, MSDs
are diffusive with $\langle [\Delta \tilde{x}(t)]^2 \rangle =2 D_a
t$, $\langle [\Delta \tilde{y}(t)]^2 \rangle = 2 D_b t$, $\langle
[\Delta x(t)]^2 \rangle = \langle [\Delta y(t)]^2 \rangle=(D_a
+D_b)t\equiv 2 \overline{D} t$ and $\langle( \Delta \theta(t) )^2
\rangle = 2 D_\theta t$.

\section{Results and Discussion}

\begin{figure}
  \centering
  \includegraphics[width=\columnwidth]{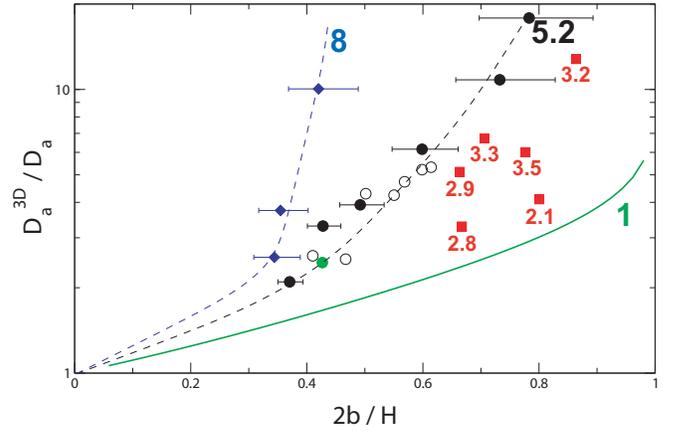}
  \caption{(color online) Ratio of theoretical 3D diffusion coefficient
  \cite{Happel91} along the ellipsoid long axis, $D_a^{3D}$, to the measured diffusion coefficient, $D_a$, for ellipsoids confined at $2b/H$.  Diamonds: $2.4\times 0.3\times 0.3~\mu$m$^3$ ($\phi=8$)
ellipsoids; Circles: $3.3\times 0.635 \times 0.635~\mu$m$^3$
($\phi=5.2$) ellipsoids; Solid circles: From samples with no added
salt; Open circles: From samples with 2 mM added salt; Green solid
circle: From sample with BSA (bovine serum albumin) covered glass
surfaces; Squares: All other samples - lower aspect ratio spheroids
with particle aspect ratios labeled below each data point.  The
accuracy of  $2b/H$ for these measurements is similar to other
samples. Dashed curves: Guides for the eye. Solid curve ($\phi=1$):
Replot of the numerical prediction in Fig.~1 of
Ref.~\cite{Bhattacharya05} for a sphere strictly in the $H/2$
mid-plane.}
  \label{fig:Da}
\end{figure}

\begin{figure}
  \centering
  \includegraphics[width=\columnwidth]{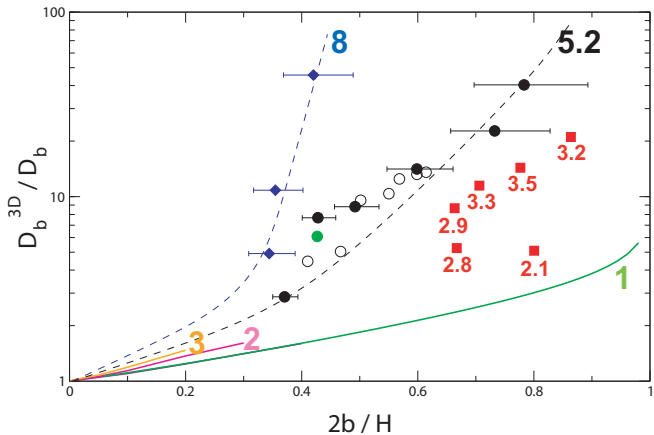}
  \caption{(color online) Ratio of theoretical 3D diffusion coefficient
  \cite{Happel91} along the ellipsoid short axis, $D_b^{3D}$, to the measured diffusion coefficient, $D_b$, for ellipsoids confined at $2b/H$. Symbols are the same as those in
Fig.~\ref{fig:Da}. Solid curves from left to right: theoretical weak
confinement predictions \cite{Happel91} for aspect ratios $a/b=3,2$,
and numerical result \cite{Bhattacharya05} for aspect ration $a/b=1$
ranging over both weak and strong confinement regimes. }
  \label{fig:Db}
\end{figure}

\begin{figure}
  \centering
  \includegraphics[width=\columnwidth]{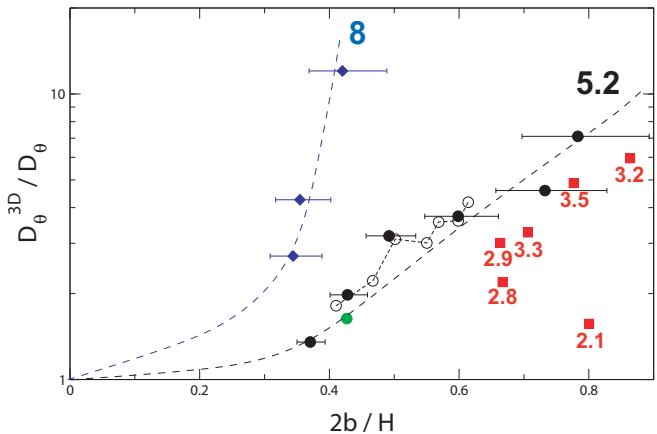}
  \caption{(color online) Ratio of theoretical rotational diffusion coefficient
\cite{Perrin34,Koenig75} $D_{\theta}^{3D}$ to the measured diffusion
coefficient, $D_{\theta}$, for ellipsoids confined at $2b/H$.
Symbols are the same as those in Fig.~\ref{fig:Da}. }
  \label{fig:Dangle}
\end{figure}

We repeated the experiments described above for different ellipsoids
under different confinement conditions. From the slopes of their
MSDs, we obtain $D_a$, $D_b$ and $D_{\theta}$ of different particles
as a function of confinement condition as shown in
Figs.~\ref{fig:Da}, ~\ref{fig:Db} and ~\ref{fig:Dangle},
respectively.  Specifically, the normalized quantities, $
D_i^{3D}/D_i = \gamma_i / \gamma^{3D}_i$, for $i = a, b, \theta$,
are plotted as a function of increasing confinement, $2b/H$.  Here
the 3D normalization constants $ D_i^{3D}$ (alternatively,
$\gamma^{3D}_i$), are calculated from Eqs.~\ref{eq:Stokes} and
\ref{eq:G}.

Notice that $2b/H=0$ corresponds to the 3D limit wherein
$D^{3D}/D=1$.  As expected, hydrodynamic drag increased and the
diffusion coefficients correspondingly decrease as the confinement
becomes stronger. The larger positive slopes exhibited by the more
needle-like spheroids are indicative of motions more sensitive to
confinement. Furthermore, the slopes of the same ellipsoids in
Fig.~\ref{fig:Db} are larger than those in Fig.~\ref{fig:Da},
indicating that diffusion along the ellipsoid short axis is more
strongly affected by the confinement than diffusion along the long
axis. Limited comparisons with all available analytical and
numerical predictions (i.e. the solid curves in Figs.~\ref{fig:Da},
~\ref{fig:Db}), suggest that our data exhibit the generally expected
trends with increasing confinement. Note that solid curves of
analytical and numerical predictions in Figs.~\ref{fig:Da},
~\ref{fig:Db} are for particles forced in the $z=H/2$ mid-plane. In
real experiment, the measured drag is an average at different $z$.
In our experiments, there are no detectable interference color
changes at the centers of ellipsoids. Consequently $z$-fluctuations
are less than $\sim 50~\unit{nm} \sim H/20$. In contrast, numerical
results in Ref.~\cite{Bhattacharya05} show that the drag of a sphere
at $H/3$ is very close ($< 10$\%) to the drag at $H/2$. Thus the
$z$-fluctuations of our ellipsoids should have negligible effects on
particle drags.

\begin{figure}[!t]
  \centering
  \includegraphics[width=\columnwidth]{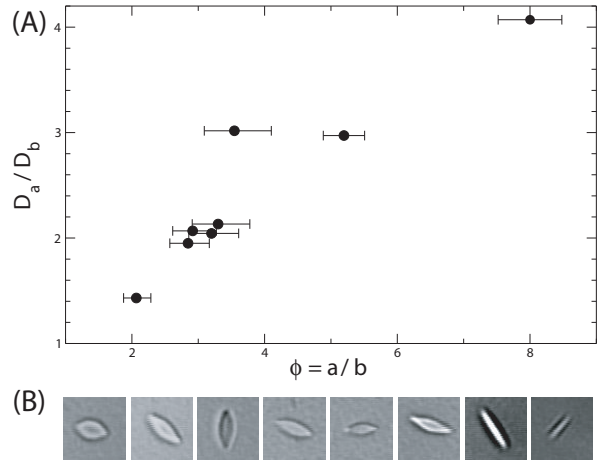}
  \caption{(A) Ratio of diffusion coefficients $D_a/D_b$ versus aspect ratio $\phi=a/b$. Seven of eight data points were taken for $2b/H \simeq 0.8$; the data point at $a/b=8$ was taken with $2b/H \simeq 0.4$.  Solid curve: Theoretical result for $D_a/D_b$ versus aspect ratio $phi=$$a/b$ in 3D. (B) Microscope images of the corresponding eight particles as a function of  aspect ratio (increasing from left to right). The dimensions of each particle from left to right: $[2a, 2b] = [2.48,1.2]$, $[2.96,1.04]$, $[2.8,0.96]$, $[2.88,0.9]$, $[2.64,0.8]$, $[3.12,0.88]$, $[3.3,0.635]$, $[2.4,0.3]~\mu$m.}
  \label{fig:Dratioab}
\end{figure}

Another question that our data holds potential to explore concerns
the effect of electric double layers on ellipsoid diffusion.  The
electric double layers around charged particles in suspension
increase their hydrodynamic diameter and slow down diffusion,
especially rotational diffusion because rotational drag is
proportional to the volume rather than the length of the ellipsoid.
This effect in rotational diffusion has been observed with
depolarized dynamic light scattering in the regime where ionic
concentration was low and spheroids small \cite{Matsuoka96}.   In
our systems such effects are expected to be small due to the high
ionic strength of the suspension. As can be seen in
Figs.~\ref{fig:Da}, \ref{fig:Db} and \ref{fig:Dangle}, diffusion
coefficients are indistinguishable for the samples with 2 mM added
salt and no added salt. The 6.9 nm screening layer of 2 mM solution
lowers 3D diffusion coefficients by less than 2\%.

\begin{figure}[!t]
  \centering
  \includegraphics[width=0.8\columnwidth]{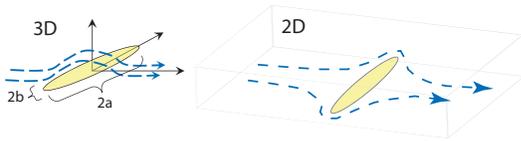}
  \caption{Schematic of fluid flow around an ellipsoid perpendicular to the flow.
For the drag along short axes, fluid can flow around $2b$ in 3D, but
has to flow around long axis $2a$ in 2D.}
  \label{fig:flowfield}
\end{figure}

Finally, Figure~\ref{fig:Dratioab} shows the impact of aspect ratio
on the ratio $D_a/D_b$. Here it is evident that diffusion in
quasi-2D is quite different from diffusion in 3D. For 3D, $D_a/D_b$
asympotes to 2 at large aspect ratio, as shown by the solid
theoretical curve.  For quasi-2D, $D_a/D_b$, on the other hand,
grows very rapidly with increasing aspect ratio Since we expect the
stick boundary condition to hold in this system, the observation
that $D_a/D_b=\gamma_b/\gamma_a>2$ should be purely due to
confinement. A schematic to qualitatively capture this basic effect
is given in Fig.~\ref{fig:flowfield}. Imagine the fluid flowing past
the ellipsoid.  In 3D, the fluid flow pathways will be displaced by
distances of order $2b$ in order to `go around' the ellipsoid.  This
fluid flow displacement is the approximately the same, whether the
spheroid is orientated either parallel or perpendicular to the flow,
and therefore $\gamma_a$ and $\gamma_b$ are comparable. In 2D,
however, the fluid flow pathway displacement is approximately $2b$
(or $2a$) when the spheroid is oriented parallel (perpendicular) to
the flow, so that $\gamma_b/\gamma_a$ diverges with $a/b$. This
qualitative picture also explains our observation that diffusion
along the ellipsoid short axes is more strongly affected by the
confinement than diffusion along the long axis.  Finally, we note
that in quasi-2D confinement, $D_a/D_b$ increases with increasing
aspect ratio and should eventually saturate ~\cite{Bhattacharya05}
at a value much larger than two, because some fluid will `leak'
between the particle and walls.

In summary, we have found that the anisotropic drag coefficients for
ellipsoid diffusion substantially increase when the ellipsoids are
strongly confined, especially along the short axes.  In the future
many questions remain about these systems will be exciting to
explore, including the effects of neighboring ellipsoids and the
effects of other confinement geometries. For example, quasi-1D
confinement of an ellipsoid will align the ellipsoid along the
diffusion direction. This effect may compensate the drag from
boundaries and lead to an optimal diameter for ellipsoid diffusion
speed in a quasi-1D cylinder.

\section{Acknowledgement}
We thank Jerzy Blawzdziewicz for providing us with the theoretical
curves at $\phi=1$ in Fig.~\ref{fig:Da} and \ref{fig:Db}. We also
thank Tom Lubensky for useful conversations.  This work was
supported by the NSF-MRSEC grant \#DMR-0520020 and partially by the
NSF grant \#DMR-0804881.


\end{document}